\documentstyle[12pt,psfig]{article}
\begin{document}

\begin{center}
{\Large\bf Accelerating  Universe from Extra Spatial Dimension}\\[20 mm]
S.~Chatterjee \footnote{Permanent Address :
 New Alipore College, Kolkata  700053, India\\[1mm] Correspondence to: S. Chatterjee,
 email : sujit@juphys.ernet.in} and A. Banerjee\\

 Relativity and Cosmology Centre,
Jadavpur University, Kolkata - 700032, India\\

                  and\\

Y.~Z.~Zhang\\

Institute of Theoretical Physics, Beijing 100080, China

\end{center}

\begin{abstract}

We present a simple higher dimensional FRW type of model where the
acceleration is apparently caused by the presence of the extra
dimensions. Assuming an ansatz in the form of the deceleration
parameter we get a class of solutions some of which shows the
desirable feature of dimensional reduction as well as
 reasonably good physical properties of matter. Interestingly we do not
  have to invoke an extraneous scalar field or a cosmological constant
 to account for this acceleration. One argues that the terms containing
 the higher dimensional metric coefficients produces an extra
 negative pressure that apparently drives the inflation of the
 4D space with an accelerating phase.
 It is further found that in line with the physical requirements our
  model admits of a decelerating phase in the early era along with an
   accelerating phase at present.Further the models asymptotically mimic
 a steady state type of universe although it starts from a big type of
  singularity.Correspondence to Wesson's \emph{induced matter theory} is
 also briefly discussed and in line with it it is argued that the terms 
containing the higher dimensional metric coefficients apparently creates a negative 
pressure which drives the inflation of the 3-space with an accelerating phase. 
\end{abstract}


\bigskip
\section*{1.INTRODUCTION }
One of the greatest challenges of modern cosmology is
understanding the nature of the observed late-time acceleration of
the universe. Recent measurements of type Ia Supernovae (SNIa)
\cite{permutter} at redshifts $z\sim 1$ and also the observational
results coming from the cosmic microwave background radiation
along with the Maxima \cite{cbn} and Boomerang data \cite{babli}
indicate that the expansion of the present universe is
accelerated. In fact the present day results show that the
supernovae look fainter than that expected from the luminosity
redshift relationship in a decelerating universe. One is tempted
to believe that in the universe there exists an important matter
component which , in its most simple description , has the
characteristic of a cosmological constant i.e.,a vacuum energy
density which contributes to a large component of a negative
pressure. However, the inability of the particle physicists to
compute the energy of the quantum vacuum~- ~contributions from
well understood physics amount to $10^{55}$ times the critical
density - casts a dark shadow on the feasibility or otherwise of
the presence of the constant. A rather important issue is the
coincidence problem: dark energy seems to start dominating the
energy budget and accelerating the expansion of the universe, just
around the present time. To circumvent this difficulty and a host
others an evolving cosmological constant or a time dependent
scalar field termed \emph{quintessence} with an appropriate self
interaction potential $ V(\phi)$ are introduced in the theory.The
apparent proliferation of quintessence dominated models is due to
the fact that it is trivial\cite{pd} to chose the 'appropriate`
potential $ V(\phi)$ such that one can always explain the
observations for any given pair of scale factor and matter
density. A number of other candidates have also been proposed:
Brans-Dicke type of scalar fields \cite{narayan} and a network of
frustrated topological defects, to name a few. As the large scale
structure and CMBR observations referred to earlier suggest that
the universe is spatially flat,with matter density equal to about
one third of the critical density the idea of a dark energy
component is gaining momentum. While these and other models have
some motivation and also attractive features none of them are
compelling because of the fact that the cosmological constant or
an evolving quintessence energy require extremely small number to
fit the data:~the present value of the energy density ,
$\rho_c\sim 10^{-12}eV$, and in the case of quintessence, the tiny
mass smaller than the current Hubble parameter $m_{Q}<H_0 \sim
10^{-33}eV$ and sub-gravitational couplings to visible matter to
satisfy fifth force constraints \cite{carrol}.In view of the fact
that the observational data coming from the supernova studies
probe length scales $ l\sim H_{0}^{-1}\sim 10^{3}$ Mpc , which are
so far inaccessible to any particle physics experiments , various
alternative proposals \cite{wilt} have cropped up time to time to
account for the current accelerated expansion without assuming any
form of dark energy.They include phantom energy\cite{phantom},
certain modification of GR \cite{gr}, Chaplygin gas as also its
variable form \cite{zhang}. Very relevant to mention that
Vishwakarma \cite{vis} in a series of works took a completely
opposite view to argue that it is possible to explain the dimming
of supernovae within the framework of of the conventional
decelerating model itself. Another line of argument is put forward
by to suggest \cite{carol} that the radiation coming from the SNe
are partially absorbed due to encounter by an obstacle. But
observationally there is no frequency dependence of the dimming of
the supernovae and so the mechanism must be very achromatic which,
in turn, seems to rule out a medium of material particles.Because
they may absorb light in the optical spectrum but will re-emit it
in IR, affecting the CMBR in unacceptable ways. The dimming of the
supernova is also sought to be explained on the basis of the so
called \emph{flavor oscillation} ( see reference 7) so that
 light travelling through inter galactic magnetic fields can partially
 be converted into axions, and evade detection on earth. So the
 source would appear fainter and hence more distant than it is actually is
 although the Universe is not accelerating.In view of what has been stated
  above we have thought it worthwhile to incorporate the phenomenon
   of accelerating universe in the framework of higher dimensional
   spacetime itself. Here we have examined a scenario in multidimensional
spacetime where the accelerated expansion of the universe at the
current epoch may be made possible without forcing ourselves to
invoke any time dependent extraneous scalar field or vacuum
energy.We have here taken a five dimensional spatially  flat,
homogeneous spacetime with perfect fluid as matter field and
assuming a specific form of
 the \emph{deceleration parameter} we have been able to show that
the universe decelerates at the early era ( a good news for
structure formation) and after a certain instant starts
accelerating in conformity with the present day observations.
Higher dimensional spacetime is now an active field of activity in
both general relativity and particle physics in its attempts to
unify gravity with all other forces of nature  \cite{appel}.
 These theories include kaluza-klein , induced matter, super string,
supergravity and string. In these (4+d) dimensional models the
d-spacelike dimensions are generally spontaneously compactified
and the symmetries of this space appear as gauge symmetries of the
 of the effective 4D theory. At present these extra dimensions are
 not observed presumably because with time they shrink to an
  unobservably small length, say plankian.However standard cosmology
  indicates that the scale factor for the extra dimensions at some
   epoch in the past could have been comparable with or even larger
  than,that of the usual three dimensional space.Renewed interests to
   these models also stem from their recent applications to brane-
  cosmology.Our paper is organised as follows: After Introduction
  in section 1 we discussed the mathematical formalism and its
   implications in section 2 where we have, for simplicity, discussed
   two solutions from the class of solutions obtained. Interesting to
   point out that depending on the choice of the arbitrary constants
   our solutions exhibit the desirable property of dimensional
   reduction.Further our cosmology assumes a steady state type
   behaviour with time although preserving the big bang type
   singularity. We argue that the late acceleration may be explained
   on the basis of Wesson`s induced matter theory according to which
   the extra dimension creates an effective 4D pressure, which in the
   present case is incidentally negative.
    The paper ends with a short discussion in section 3.

\section*{2.MATHEMATICAL FORMALISM}
We here discuss a spatially flat 5D homogeneous cosmological model with the topology
$M^1 \times R^3 \times S^1$ where $S^1$ is taken in the form of a circle such that
\begin{equation}
ds^2  = dt^2 - R^{2}(t) \left(dr^2 + r^2 d\theta^2 + r^2 \sin^2 \theta d \phi^2 \right)
 - \Phi^{2}(t) dy^2
\end{equation}
where $R(t)$ is the scale factor for the 3D space and $\Phi(t)$,
that for the extra dimension and y is the fifth dimensional
coordinate. \noindent The metric (1) admits a number of isometries.
For the case $R^{1} X S^{3} X S^{1}$ the symmetry group of the group
of spatial section is $O(4) X O(2)$.The stress tensor whose form will
be dictated by Einstein's equations must have the same invarience.
Therefore the energy momentum tensor may be written as follows\cite{salam}\\

$ T_{00}= \rho(t), T_{ij} = p(t)g_{ij}, T_{55} = p_{5}g_{55}$\\

where the rest of the components vanish.
The conservation of the energy-momentum tensor $ T^{ab};b = 0$ yields

\begin{equation}
\dot{\rho} + ( p + \rho)3\frac{\dot{R}}{R} + (p_{5}+\rho)\frac{\dot{\Phi}}{\Phi}
= 0
\end{equation}
where $p_{5}$ is the pressure in the fifth dimension while that
for the 3D space is isotropic and is given by $p$\\
The independent field equations for the metric (1) and matter
field (2) are given by
\begin{eqnarray}
3 \frac{\dot{R}^2}{R^2}  + 3\frac{\dot{R}}{R} \frac{\dot{\Phi}}{\Phi}  &=&  \rho \\
2 \frac{\ddot{R}}{R}  + \frac{\dot{R}^2}{R^2}  + 2
\frac{\dot{R}}{R}\frac{\dot{\Phi}}{\Phi} +
\frac{\ddot{\Phi}}{\Phi}  &=& - p \\
3\frac{\ddot{R}}{R} + 3 \frac{\dot{R}^2}{R^2} &=& -p_{5}
\end{eqnarray}

As we have five unknowns ( R, $\Phi$, $\rho$ , p and $p_{5}$ ) with three
independent equations we are at liberty to choose two connecting
relations among them.We assume that $p = p_{5}$ so the pressure is
isotropic including the extra space .From this condition we get
from the field equations
\begin{equation}
\ddot{\Phi}+ 2\frac{\dot{\Phi}\dot{R}}{R} -
\Phi(\frac{\ddot{R}}{R} + 2 \frac{\dot{R^{2}}}{R^{2}}) = 0
\end{equation}

A little inspection shows that $ \Phi = R$ is a solution of this
equation , which makes the spacetime isotropic. For a more general
solution we substitute  $ \Phi = Ru(t)$ such that
\begin{equation}
R\ddot{u}+ 4 \dot{R}\dot{u}  = 0
\end{equation}
or $ \dot{u} = \frac{\beta}{R^{4}}$ where $\beta$ is an arbitrary
constant.As we have another choice we make the ansatz that the
 \emph{deceleration parameter},
\begin{equation}
q = -\frac{R\ddot{R}}{\dot{R}^{2}}
\end{equation}
is given by
\begin{equation}
q = \frac{a-R^{m}}{b+R^{m}}
\end{equation}
where $m$ is any arbitrary constant. In taking this particular
form of the expression we are guided by the consideration that the
deceleration parameter should be amenable to both decelerated
expansion at the early era and accelerated expansion at the
current era to account for both structure formation as well as for
interpretation of the current data from the supernova.Naturally
both the constants a and b should be positive. The expression
\begin{equation}
\frac{\ddot{R}}{\dot{R}} =
-\frac{a-R^{m}}{b+R^{m}}(\frac{\dot{R}}{R})
\end{equation}
offers a first integral as
\begin{equation}
\dot{R} = ( R^{m}+ b )^{2} R^{\frac{2-m}{2}}
\end{equation}
A little inspection shows that the expression $ R=sinh^{2/m}\omega
t$ , subject to the restrictions that $ b=1$ and $ a = \frac{m -
2}{2}$ will be a solution of this equation. For convenience we
take $2/m = n$ , which finally reduces q to
\begin{equation}
q = \frac{(1 - n ) - n~sinh^{2}\omega{t}}{n~cosh^{2}\omega{t}}
\end{equation}
The nature of variation of, q depends critically on the value of
the exponent n, which in turn, determines the scale factor R. For
$n > 1$ we get only accelerating model.However for $ n < 1$ the
desirable phenomenon of flip occurs. An initially decelerating
model starts accelerating after a certain instant. But it is not
obvious from our analysis at what value of redshift this flip
occurs. We here consider only two cases corresponding to $ n =~
1/2~~ and ~~1/4$. As we are considering an expanding model any
negative value of n is ruled out. Skipping intermediate
mathematical steps for economy of space we give the final
expressions as follows :

 \textbf{CASE I} ( n= 1/2 )\\
In this case we get
\begin{eqnarray}
\Phi &=& sinh^{\frac{1}{2}}\omega t(\gamma - \beta coth\omega t )\\
R &=& R_{0}sinh^{\frac{1}{2}}\omega t
\end{eqnarray}
With the above form of the metric coefficients the matter field
reducet to
\begin{eqnarray}
p &=& p_{5} = - 3\omega^{2}/2\\
\rho &=& \frac{3}{2}\frac{cosh^{2}\omega t}{sinh^{2}\omega t} +
\frac{3\beta cosh \omega t}{2sinh^{3}\omega t(\gamma - \beta coth t )}
\end{eqnarray}
Interestingly as $t \sim 0$ , $\rho = \rho_{0} =
\frac{3\omega^{2}}{2}$ and as $t \sim \infty$, $\rho_{\infty}=
\frac{3\omega^{2}}{2}$ So asymptotically the mass density tends to
assume a constant value and the cosmology mimics a steady state
type of behaviour though not exactly following the type advocated
by Bondi and Gold or Hoyle and Narliker. With time the density
assumes a constant value as $ \rho_{\infty} = 3\omega^{2}/2$.
Further as $t \sim 0 $, $R \sim t^{1/2}$ and $ A \sim t^{-1/2}$ So
at the early era the spacetime resembles the well known form given
by Chodos and Detweiler \cite{cd}. The temporal behaviour of the
model depends critically on the initial conditions. If the
arbitrary constant $\gamma$ is made zero the extra dimension
starts from an infinite extension, shrinks to a minimum and then
expands again indefinitely so that there is no dimensional
reduction in this case However the large extra dimensions in the
theory are not that much out of favour these days as it attempts
to address the well known hierarchical problem in quantum field
theory.Further with time the model isotropises in this case and
the 4D volume expands.On the other hand with non vanishing
$\gamma$ the extra dimension contracts and ultimately vanishes
exhibiting the desirable feature of dimensional reduction. In this
case the model again becomes singular Theorists try to save the
situation via assuming varied stabilising mechanisms  \cite{zhuk}
like quantum gravity, Casimir effect etc. so that they produce a
sort of repulsive potential to halt the shrinkage at a very small
constant value,say planckian length. At this stage the extra
dimensions essentially decouple as all its derivatives vanish in
the field equations. The cosmology now enters the standard 4D
phase following the FRW model without having any reference to the
extra dimensions.\\A serious shortcoming of this analysis is that
there is no dynamical evolution in the expression for pressure. We
shall shortly see that this defect is overcome in our next
section.

\textbf{CASE II} ( n= 1/4 )\\

Skipping the intermediate mathematical steps like case I we write
the final results as
\begin{eqnarray}
R &=& R_{0}sinh^{\frac{1}{4}}\omega t \\
\Phi& = & sinh^{\frac{1}{4}}\omega t (\beta~ln ~tanh\frac{\omega
t}{2} + \gamma )\end{eqnarray}
 With the above values of the R and$\Phi$we get the following expressions
 for the pressure matter density as
 \begin{eqnarray}
 p &=& p_{5} = - \frac{3\omega^{2}}{8 sinh^{2} \omega t} ( sinh^{2}\omega t - 1)\\
 \rho &=& \frac{3 \omega^{2}}{8}\frac{cosh \omega t}{sinh^{2} \omega t} ( cosh\omega t
 + \frac{2 \beta}{\beta ~ln~tanh\omega t + \gamma})
 \end{eqnarray}

As commented earlier this model does not suffer from the
disqualification of a constant pressure. Here both pressure and
mass density are evolving and start from an infinite value as a
big bang singularity.But it has not escaped our notice that both
the physical quantities assume steady values asymptotically at
$\frac{3\omega^{2}}{8}$. So unlike the big bang type it resembles
more a steady state type cosmology.But the pressure changes
signature from positive to negative at $ sinh\omega t =1$. However
it is not difficult to to explain the asymptotically steady value
of the matter field because a little algebra shows that with time
the 4D volume,
\begin{equation}
V =  R^{3}\Phi = sinh\omega t~ ( \beta~ ln~tanh\omega t + \gamma )
\end{equation} stabilises at some finite value.Another striking
difference from the earlier case is that here both the scales
start from zero and depending on the signature of the arbitrary
constant $\gamma$ the fifth dimension either expands indefinitely
or collapses at a finite time.\\
\bigskip

\textbf{ACCELERATED UNIVERSE} \\
 As commented in our introduction this model admits of both
deceleration at the early phase and acceleration at present. The
early deceleration is physically relevant in the sense that it
allows structure formation to take place while the present day
acceleration is in conformity with the current observations. For
the general case $R = sinh^{n}\omega t $, we get for deceleration
parameter
 $q = - \frac{\ddot{R}R}{\dot{R^{2}}}= - \frac{ncosh^{2}\omega t - 1}{n cosh^{2}\omega t}$.
 Thus for $n \geq 1$, $q < 0$. Hence always accelerating.\\

 For case I, ~~ $n=\frac{1}{2}$ and $q = \frac{1 - sinh^{2}\omega t}{1 + sinh^{2}\omega t}$.
 Let $ sinh \omega t_{c} = 1$. Hence for $t < t_{c}$ ,~~ $q > 0$ (deceleration) and for
 $t > t_{c}$ ,~~ $q < 0$ (acceleration).\\
 \\
 For case II, ~~ $n= \frac{1}{4}$ ~~and ~~
 $q = \frac{3- sinh^{2}\omega t}{ cosh^{2}\omega t} $.
 Let $ sinh \omega t_{c} =\\\sqrt{3}$. Hence for $t < t_{c}$ ,~~ $q > 0$ (deceleration) and for
 $t > t_{c}$ ,~~ $q < 0$ (acceleration).\\

Hence acceleration starts later in the second case.\\

Before ending the section a final remark may be in order.
For example if we take the case II it follows that the acceleration starts 
when $sinh~wt > \sqrt{3}$. On the other hand the pressure starts becoming
negative at $sinh~wt>~1$. Evidently acceleration of the universe starts
later than the instant when the pressure becomes negative. One can
argue that the fact that the pressure becomes negative does not
guarntee the present acceleration of the universe. For this to happen
it has to dominate long enough to overcome the gravitational
attraction produced by ordinary matter.

\textbf{ WESSON'S FORMALISM}\\

At this stage one may digress a little and call attention to the
\emph{Induced Matter Theory} recently developed and formulated by
Wesson  \cite{wessson}according to which it is possible to
interpret most properties of matter as a result of 5D Riemannian
geometry.Accordingly the 5D field equations for the apparent
vacuum for Einstein tensor are given by
\begin{equation}
G_{A B} =0
\end{equation}

By contrast the 4D equations for Einstein`s equations are given by
\begin{equation}
G_{ij}=~T_{ij}
\end{equation}
where A, B run from 0 to 4 where as i, j from 0 to 3. The central
idea of the induced matter theory ( see Wesson, p.42 for more
details) is that the equations (23) are a subset of (22) with an
effective or induced energy momentum tensor $T_{ij}$ which contain
the classical properties of matter.It follows from the theorem of
Cambell that any analytic N dimensional Riemannian manifold can be
locally embedded in an ( N+1)D Ricci-flat Riemannian manifold
\cite{tavakol}. Though not exactly similar ( we are here not
dealing with a 5D vacuum) we can collect the terms containing the
fifth dimension from the expression of pressure to call them
$p_{\phi}$ such that from the equation (4 ) it follows that
\begin{equation}
p_{\phi} = - \frac{\ddot{\Phi}}{\Phi}-
2\frac{\dot{R}}{R}\frac{\dot{\Phi}}{\Phi}
\end{equation}
With our solution for $n=1/4$ the above equation yields
\begin{equation}
p_{\phi} = \frac{ 1- 3 sinh^{2}\omega t}{16~ sinh^{2}\omega t}
\end{equation}
So long as $1 > 3~sinh^{2}\omega t$  the fifth dimension induced
pressure is positive which, in turn , makes the deceleration
parameter, q is positive but soon after $p_{\phi} < 0$ and this
apparently drives the acceleration of the universe making the 3D
deceleration parameter finally negative.

\section*{3.DISCUSSION}
 While vast literature exists to address the observational fact
 of the current accelerated expansion of the universe we are not
 aware of models of similar kind in the framework of higher
 dimensional spacetime. We have here discussed a scenario in
 homogeneous 5D spacetime which admits a decelerating expansion in
 the early epoch along with an accelerated phase at present in
 line with the current observational results. The most important
 finding, in our opinion is the result that it is possible to
 achieve this acceleration without introducing any external
 quintessence-like scalar field or vacuum energy into the
 theory-the presence of the extra dimension , so to say, seems to
 cause the expansion to accelerate. One can \emph{naively} attempt to
 interpret the result as follows: It is well known that a 
higher dimensional spacetime with a Ricci-flat extra dimension
 is equivalent to an effective 4D theory with an extra
 massless scalar field,which may be instrumental in driving the
 acceleration.In this context one may also call attention to
 Wesson`s induced matter theory to argue that negative induced
  pressure is responsible for the accelerated expansion of the
  universe. Another desirable feature is the phenomenon of dimensional
  reduction so that the model finally reduces to an effective 4D one.
   This takes place in both the cases discussed here. Although
   in both the cases the cosmology starts from an initial  big bang
  type of singularity the matter field asymptotically becomes steady
   pointing to a steady state like model which is at  variance with
  the standard FRW type of models. This may be of some interests
   because we do not have to hypothesise concepts like the so-called
 `matter creation from nothing' or any type of `creation field'.
 To sum up a last remark may be in order. The first case suffers
 from the disqualification that the pressure is always a constant-
 it has no dynamical evolution. This defect is, however, not
 present in the second case. But the most serious shortcoming is
 the absence of any mechanism to achieve the stabilisation of the
 extra dimension at a very short length. It is not apparent from
 our analysis how that stabilisation work in our model. However in
 an earlier work Guendelman and Kaganovich  \cite{gk}
studied the Wheeler-Dewitt equation in presence of a negative
cosmological constant and dust. They got the interesting result
that the quantum effects do stabilise the volume of the
universe,thus providing a mechanism of quantum avoidance of the
singularity.It is also shown ( see reference 10 ) that if one one
starts with more than one extra dimension that the extra space may
generate a repulsive potential to halt the indefinite shrinkage.\\

\textbf{ACKNOWLEDGEMENT}\\

 S.C wishes to thank the Third World
Academy of Science, Trieste for Travel grant and ITP for local
hospitality during which a part of the work was done.

\end{document}